%% file: wipin.tex
\documentclass[10pt, conference, letterpaper]{IEEEtran}
\IEEEoverridecommandlockouts
\usepackage{cite}
\usepackage{amsmath,amssymb,amsfonts}
\usepackage{algorithmic}
\usepackage{graphicx}
\usepackage{textcomp}
\usepackage{xcolor}
\usepackage{subfigure}
\def\BibTeX{{\rm B\kern-.05em{\sc i\kern-.025em b}\kern-.08em
    T\kern-.1667em\lower.7ex\hbox{E}\kern-.125emX}}
\begin{document}

\title{WiPIN: Operation-free Passive Person Identification\\ Using Wi-Fi Signals
}


\author{   

\IEEEauthorblockN{Fei Wang, Jinsong Han, Feng Lin$^\dagger$\thanks{$\dagger$  Corresponding author.}, Kui Ren}
\IEEEauthorblockA{\textit{Institute of Cyberspace Research,} \textit{Zhejiang University} \\
Hangzhou, China \\
feiw2.ri@gmail.com, \{hanjinsong,flin,kuiren\}@zju.edu.cn}
}

\maketitle

\begin{abstract}
Wi-Fi signals-based person identification attracts increasing attention in the booming Internet-of-Things era mainly due to its pervasiveness and passiveness. Most previous work applies gaits extracted from WiFi distortions caused by person walking to achieve the  identification. However, to extract useful gait, a person must walk along a pre-defined path for several meters, which requires user high collaboration and increases identification time overhead, thus limiting use scenarios. Moreover, gait based work has severe shortcoming in identification performance, especially when the user volume is large. In order to eliminate above limitations, in this paper, we present an operation-free person identification system, namely WiPIN, that requires least user collaboration and achieves good performance. WiPIN is based on an entirely new insight that Wi-Fi signals would carry person body information when propagating through the body, which is potentially discriminated for person identification. Then we demonstrate the feasibility on commodity off-the-shelf Wi-Fi devices by well-designed signal pre-processing, feature extraction, and identity matching algorithms. Results show that WiPIN achieves 92\% identification accuracy over 30 users, high robustness to various experimental settings, and low identifying time overhead, i.e., less than 300ms.

\end{abstract}


\input{tex/study.tex}

\input{tex/system.tex}

\input{tex/experiment.tex}
\input{tex/related.tex}

\section{Conclusion}
In this paper, we propose WiPIN, an operation-free person identification system using Wi-Fi signals. Compared to previous work, it is much more user-friendly, time efficient, and robuster. To achieve WiPIN, we carefully design algorithms in signal processing, feature extraction, and identify matching. Besides, we prototype WiPIN in commdity off-the-shelf Wi-Fi devices and extensively evaluate WiPIN in a group of 30 subjects from various aspects. Experimental results show that WiPIN achieves competitive performance in identifying authenticated users as well as rejecting illegal users.

{\small
\bibliographystyle{IEEEtran}
\bibliography{ref}
}

\end{document}

%% file: tex/study.tex
\section{Introduction}\label{sec:introduction}

Recently, channel state information~(CSI) of Wi-Fi signals has been increasingly exploited for person identification~\cite{xin2016freesense,zeng2016wiwho,zhang2016wifi,wang2016gait,shi2017smart,lv2017device,pokkunuru2018neuralwave} due to the pervasiveness and low cost in deployment. Besides, Wi-Fi based person identification enables passive identification, i.e., high user-friendly. What's more important, unlike popular face recognition systems being vulnerable to the replay attacks \cite{smith2015face}, or fingerprint recognition struggling for spoofing attacks from 3D printed models \cite{foolphone}, it is harder to fool the Wi-Fi based person identification system because the attack needs extremely vivid imitation on user behaviors.

Among previous work, CSI monitoring during user ambulation are the most prevailing CSI-based identification approaches ~\cite{xin2016freesense,zeng2016wiwho,zhang2016wifi,wang2016gait,lv2017device,pokkunuru2018neuralwave}, which demand users to walk along pre-defined paths and record corresponding Wi-Fi CSI series simultaneously. Then the recorded CSI is either to extract specific gait metric such as walking speed for identification~\cite{wang2016gait}, or directly to learn identity classifiers with machine learning algorithms, e.g., support vector machine~\cite{xin2016freesense,zeng2016wiwho,zhang2016wifi,lv2017device} or deep neural networks~\cite{pokkunuru2018neuralwave}. However, to recognize a person's gait, one must walk along the pre-defined paths again for several meters, e.g., 2-3m in \cite{zeng2016wiwho} and 5m in \cite{xin2016freesense}, which is labor intensive and time-consuming, thus largely limits use scenarios. To overcome limitations in gait-based approaches, Shi \textit{et al.}~\cite{shi2017smart} 
propose an activity-based person identification, which is built on the finding that CSI series corresponding to user daily activities like opening micro-oven carries identity-related patterns. Though the approach in \cite{shi2017smart} requires less labor work, daily activity patterns are inferior in robustness compared to gait.

\begin{figure}[t]
    \centering
    \includegraphics[width=0.47\linewidth]{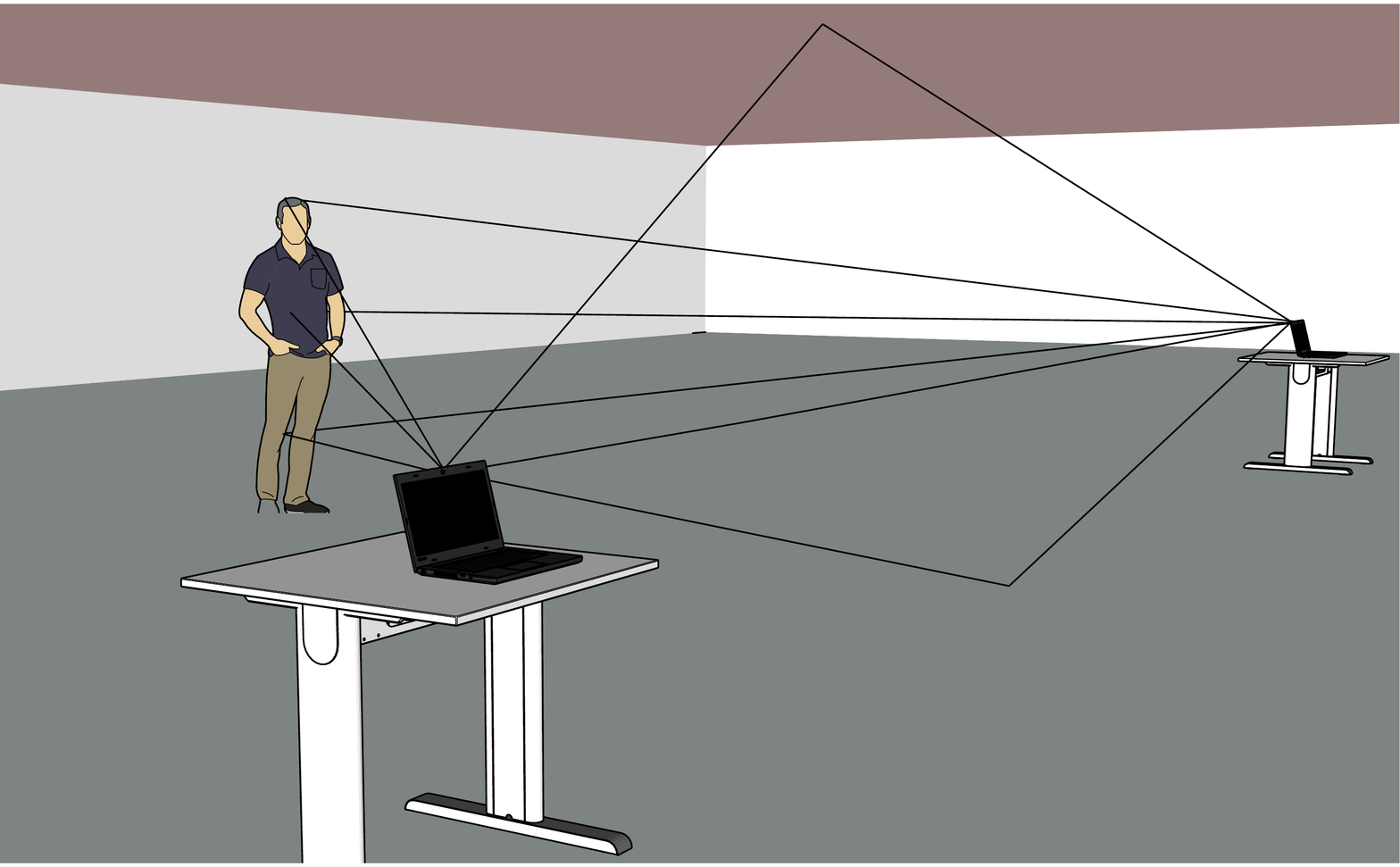}
    \includegraphics[width=0.5\linewidth]{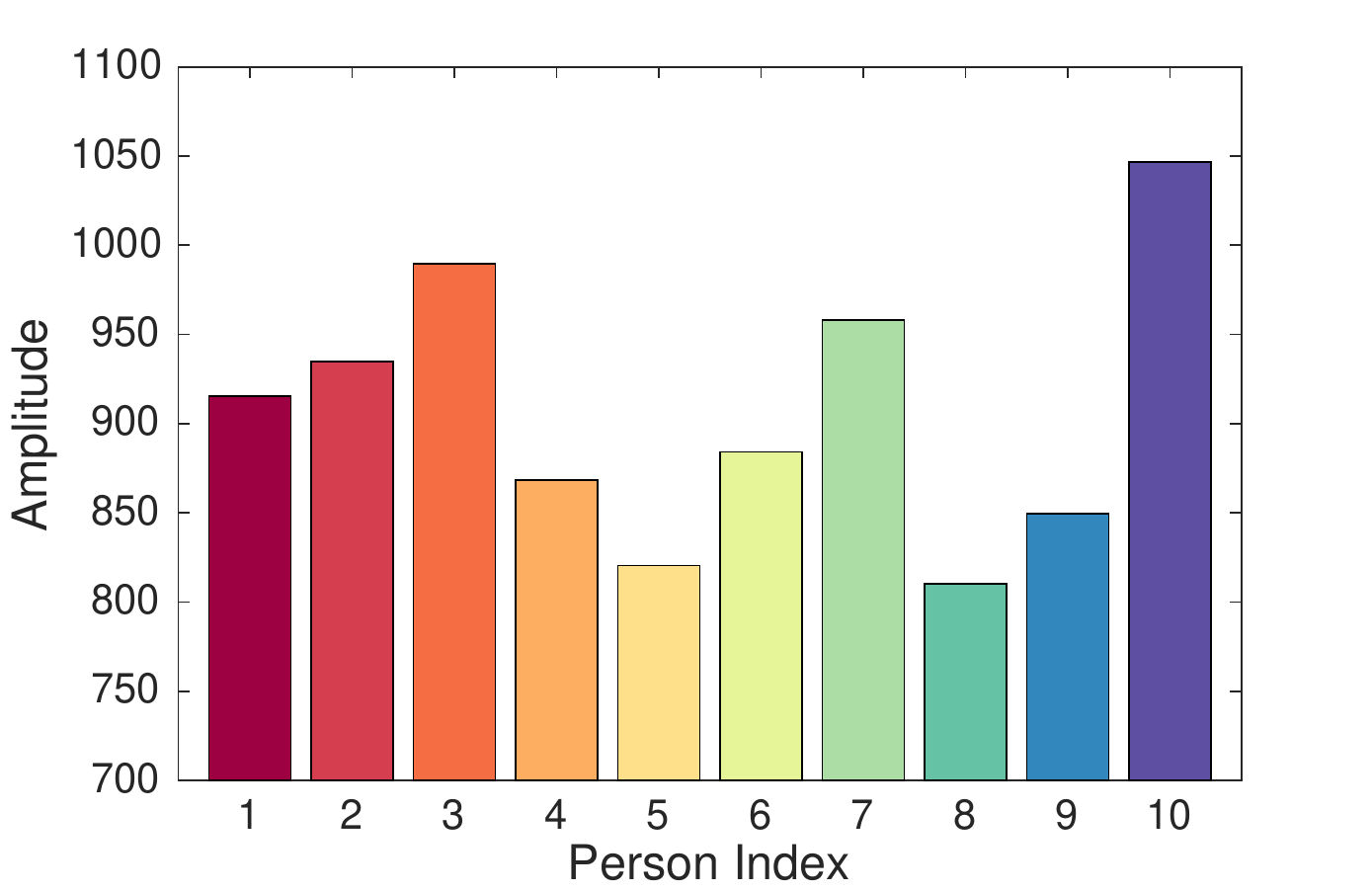}
    \caption{WiPIN rationale. When one stands in a Wi-Fi environment~(left), multi-path effect on the person body would lead to discriminated distortions in the Wi-Fi amplitude (right), which carries body information and can be well-designed for person identification.}
    \label{fig:first}
\vspace{-10pt}
\end{figure}

\begin{figure*}[t]
    \centering
    \includegraphics[width=0.95\textwidth]{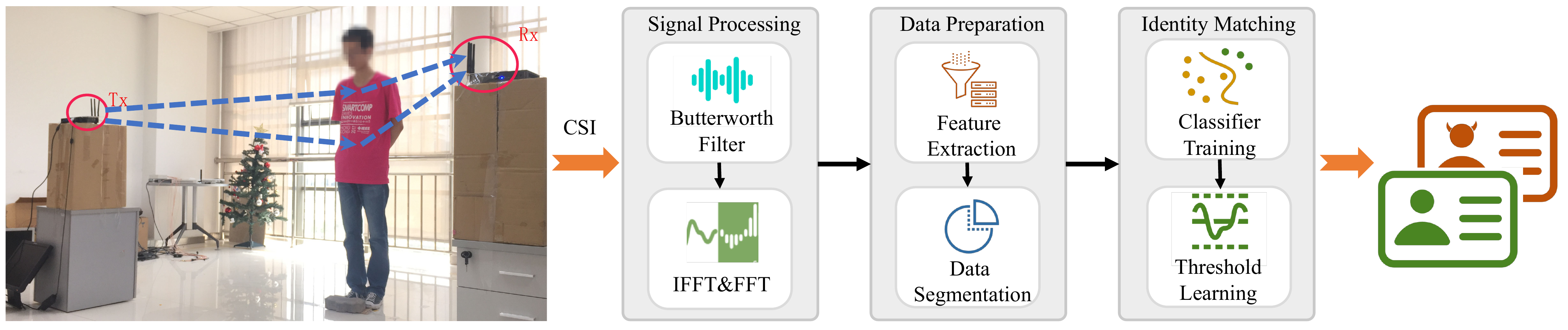}
    \caption{System framework. WiPIN is comprised of hardware and algorithms. The algorithms contain modules of signal pre-processing, data preparation, and identity matching. After deployment, WiPIN can identify legal users while reject illegal users. }
    \label{fig:system}
    \vspace{-10pt}
\end{figure*}

Previous work utilizes Wi-Fi distortions caused by user behaviors, such as walking and other daily activities embedding behavior patterns, to identify users. Moreover, we find that not only user behaviors, the presence of user can result in identity-related Wi-Fi distortions due to unique body information, e.g., body shape, body fat rate, and body muscle rate, which can also be used to identify users~(see Section~\ref{sec:study} for the study). 

With the above observation, we propose an operation-free passive person identification system, namely WiPIN. It can deeply extracts human body information from Wi-Fi distortion series caused by the user's presence, hence further identifies people with well-designed signal pre-processing and identity matching algorithms. One challenge in realizing the system is the mixture of body relevant signal and interference signals from multiple Wi-Fi propagation paths, and we tackle this problem by implementing a multi-path effect mitigation. Experimental results show that WiPIN achieves 92\% identification accuracy over 30 subjects with robustness to various experimental settings and low identifying time overhead, i.e., less than 300 ms. We summarize the main contributions of this paper as follows.

\begin{itemize}
    \item We propose WiPIN, a novel Wi-Fi signals-based passive person identification system, which has no requirement for the proactive user engagement in traditional identification systems, such as facing camera and scanning finger/iris. In addition, WiPIN is user-friendly and time-efficient in practice.
    
    \item We quantitatively study the rationale behind WiPIN, and conclude that during propagation, Wi-Fi signals are embedded with information related to human body. The intrinsic body information further used for person identification is robuster than behavior patterns. 
    
    \item WiPIN can classify authenticated users, as well as to reject illegal users that not seen before. We prototype WiPIN using commodity off-the-shelf Wi-Fi devices, and conduct extensive experiments to validate advances of WiPIN in various aspects including accuracy, robustness, time consumption, etc.   
    
\end{itemize}

\begin{figure}[b]
\vspace{-5pt}
    \centering
    \includegraphics[width=0.48\linewidth]{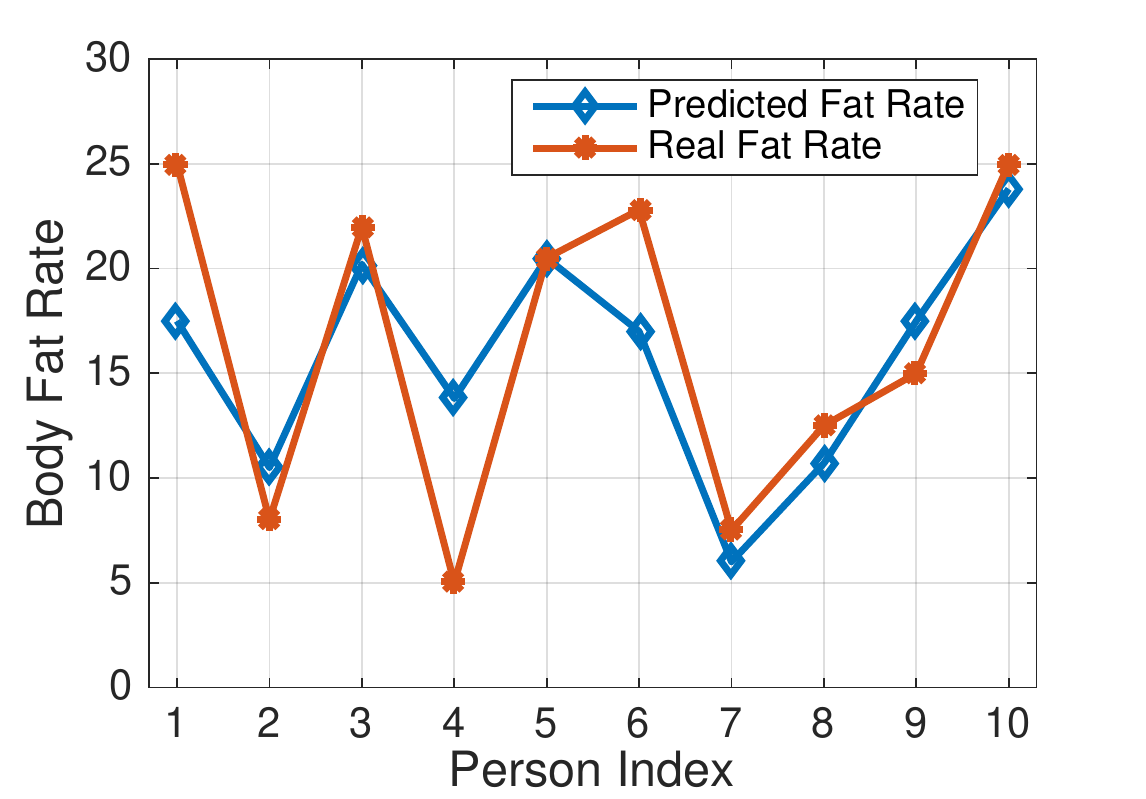}
    \includegraphics[width=0.48\linewidth]{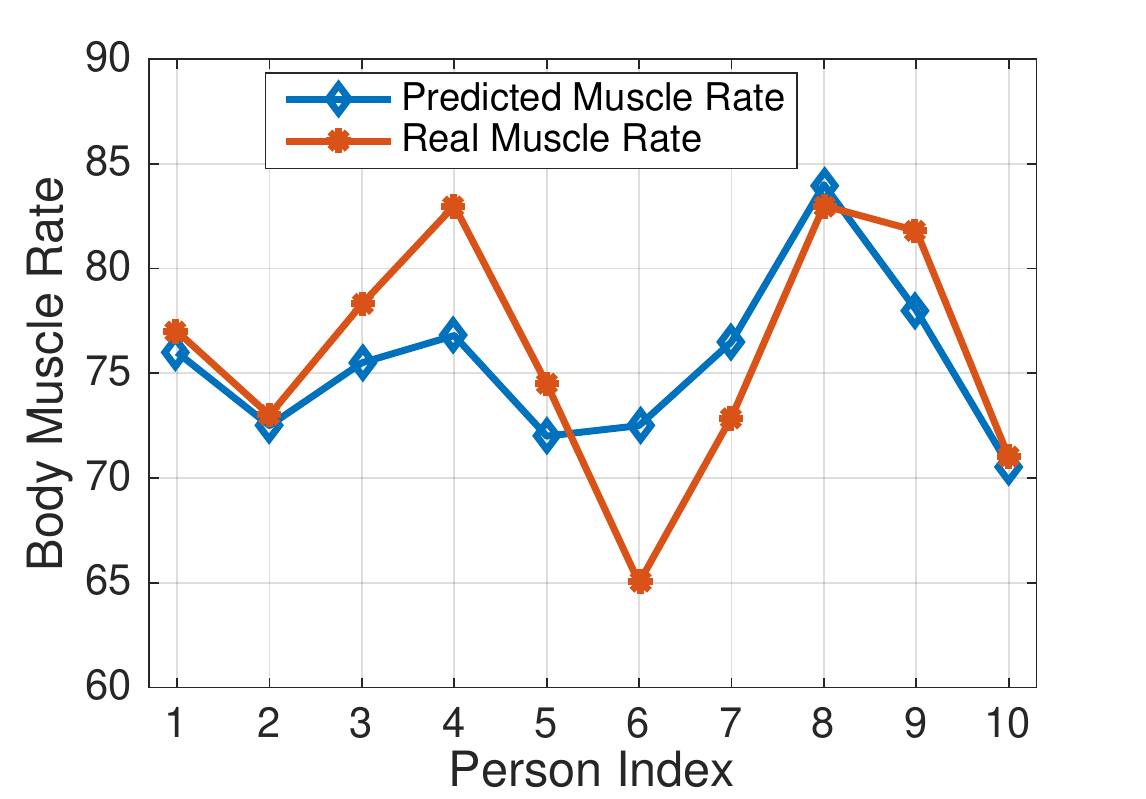}
    \caption{Information related to human body such as the fat rate and the muscle rate is embedded in the Wi-Fi distortion and can be well predicted by Wi-Fi signals with support vector regression~\cite{smola2004tutorial}.}
    \label{fig:fat}
\vspace{-10pt}
\end{figure}

\section{Rationale Study}\label{sec:study}

Past work uses Wi-Fi distortions that embedded  user behaviors patterns for person identification. Moreover, We are wondering how CSI varies if a person does no operation just standing in a Wi-Fi environment. Fig.~\ref{fig:first} illustrates our preliminary experiment results for this wondering. We recruited 10 subjects to  stand in a same position, respectively, near the Wi-Fi transmitter and receiver~(left), meanwhile the corresponding CSI series are recorded. We average the CSI amplitudes over all recorded samples and plot them in the Fig.~\ref{fig:first}~(right). Majority of the amplitudes are discriminated, which demonstrates the potentiality for person identification.

We ascribe the reason of results shown in Fig.~\ref{fig:first} to that during the propagation, Wi-Fi signals must be embedded with certain information related to human body such as body shape, body fat rate, and body muscle rate, because of the multi-path effect on the body, and the absorption and reflection effect in the body. If this guess is correct, it is possible to perform person identification using Wi-Fi signals without any user activities like walking for several meters, but only standing for a second. To further confirm the guess, we measure the body fat rate and muscle rate of these subjects by a Mi$^\circledR$ body fat scale, then train a mapping function from CSI to aforesaid rates with support vector regression (SVR)~\cite{smola2004tutorial} (our not inferring body shape is because it is hard to get the ground-truth of body shape). Fig.~\ref{fig:fat} demonstrates that CSI distortions caused by standing persons are highly relevant with body fat rate and body muscle rate.

%% file: tex/system.tex
\section{System Design}\label{sec:sys}

WiPIN can identify authenticated users while reject illegal users. Fig.~\ref{fig:system} illustrates WiPIN framework that is comprised of CSI generation hardware and person identification algorithms. In hardware-end, the transmitter~(Tx) broadcasts Wi-Fi signals, and the receiver~(Rx) records the signals. Using a Linux 802.11n CSI tool~\cite{halperin2011tool}, we can parse CSI of 30 orthogonal frequency division multiplexing (OFDM) subcarriers at 5GHz central frequency from the recorded signals. Formally, denote CSI as $H$ and suppose we record $t$ Wi-Fi samples, then $H \in C^{t\times 30} $, where $C$ is for the Complex value, which is time-serial data essentially. In this paper, we only use the amplitudes of CSI, leading to $H \in R^{t\times 30} $~($R$ for the Real value). Next we go details in the algorithm-end of WiPIN.

\subsection{Signal Processing}

\subsubsection{Noise Removal}

\begin{figure}[t]
\subfigure[Raw CSI.]{
\centering
\hspace{-0.1in}
\includegraphics[width=0.48\columnwidth]{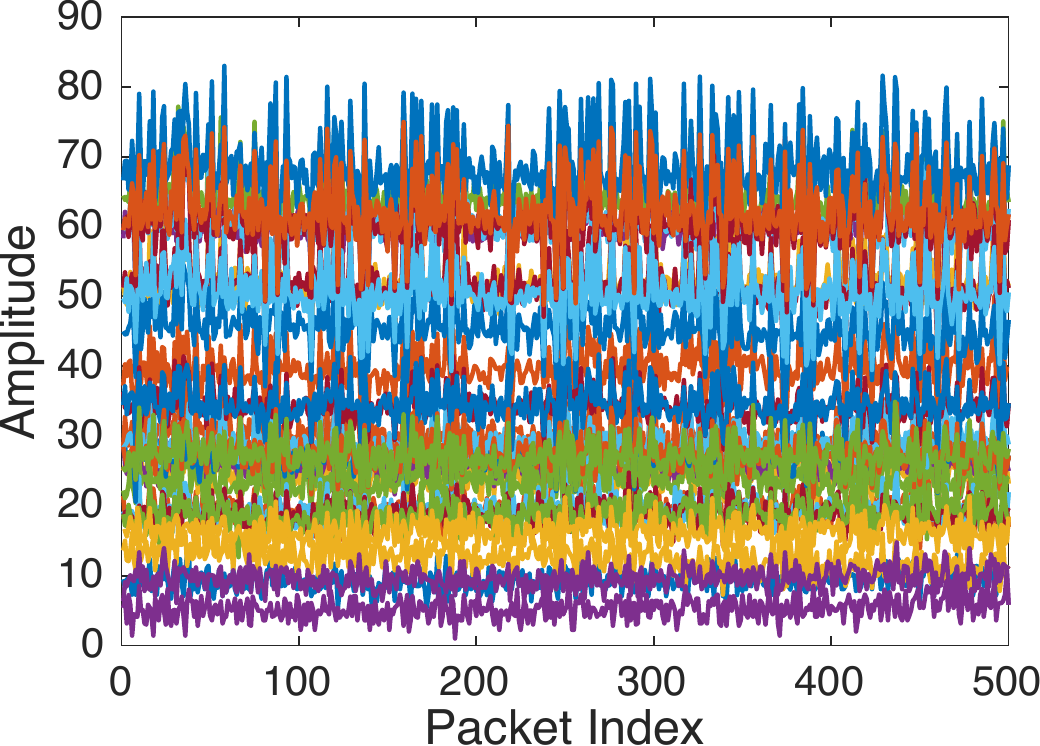}
}
\subfigure[Filtered CSI.]{
\centering
\hspace{-0.1in}
\includegraphics[width=0.48\columnwidth]{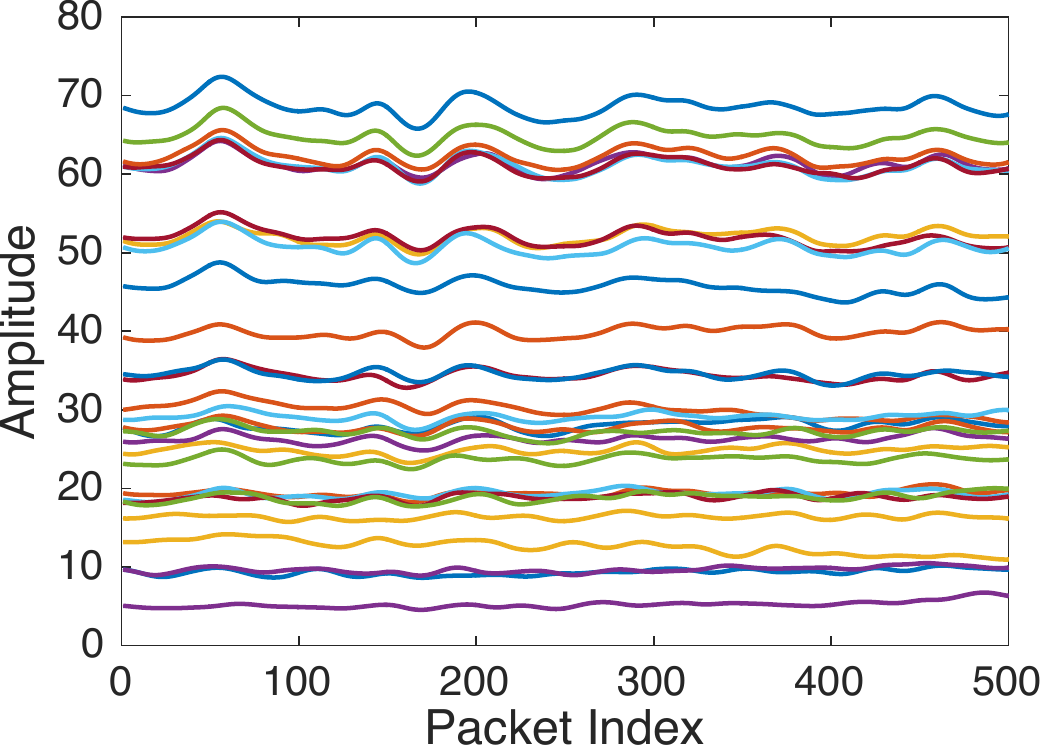}
}
\caption{Noise removal via Butterworth filter.}\label{fig:filter}
\vspace{-10pt}
\end{figure}

Due to the hardware imperfection~\cite{xie2015precise}, the sampled CSI series, i.e., $H$, have considerable noise. As an example in Fig.~\ref{fig:filter}, we plot CSI series recorded within 1 second when a subject stands as shown in Fig.~\ref{fig:system}. In the figure, each line stands for the amplitude series of one subcarrier (30 lines in all). To eliminate the high-frequency jitters, we design a low-pass Butterworth filter~\cite{selesnick1998generalized}. In particular, we experimentally set the parameter of the Butterworth filter as 5th-order with cut-off frequency of 10Hz. The filtering results are illustrated in Fig.~\ref{fig:filter} (b), which demonstrate that the Butterworth filter with above settings can significantly reduce noises in the CSI series. 


\subsubsection{Multi-path Effect Mitigation}
WiPIN uses omnidirectional antennas to broadcast and  receive Wi-Fi signals, which makes CSI the mixture of  signals from multiple propagation paths, including the line-of-sight path, the paths reflected from human body, and other reflection paths. This phenomenon is called multi-path effect~\cite{rappaport1996wireless} and can be expressed by following formula,
\begin{equation}\label{eq:multipath}
\textup{H}=\sum_{k=1}^{n}a_{k}e^{-j2{\pi}f{\tau}_{k} },
\end{equation}
where $k$ is the index of the paths, $a_{k}$ and ${\tau}_{k}$ are the power decay and time delay of the $k$-th path, respectively. 

To make CSI more relevant with person body, we aim to mitigate signals received from other paths. The bandwidth~(B) in WiPIN is 40MHz. Correspondingly, the time resolution is $\Delta t=1/B=25ns$, which yields a distance resolution of $\Delta t=c/B=7.5m$, where $c$ is the electromagnetic wave speed in the air, approximately $3\times 10^8 m/s$. In the settings of this paper~(see one example in Fig.~\ref{fig:system} leftmost), major paths reflected from person body are within $7.5m$, indicating that CSI from body paths should almost have the least time delays (within $25ns$). Thus, we first apply the inverse fast Fourier transform~(IFFT) one each sample of $H$, i.e., $h_f \in R^{1\times 30}$, to convert $h_f$ to time domain~($h_t$), keep the item in the least time delay whereas suppress the subsequent items~(divided by a large number, i.e., 1000), then convert $h_t$ back to frequency domain by fast Fourier transform~(FFT). We apply IFFT \& FFT operations on every sample of $H$, greatly mitigating multi-path effect on CSI and making CSI much more relevant to human body.

\begin{figure}[b]\label{fig:multipath}
\subfigure[CSI Time Domain.]{
\centering
\hspace{-0.1in}
\includegraphics[width=0.465\columnwidth]{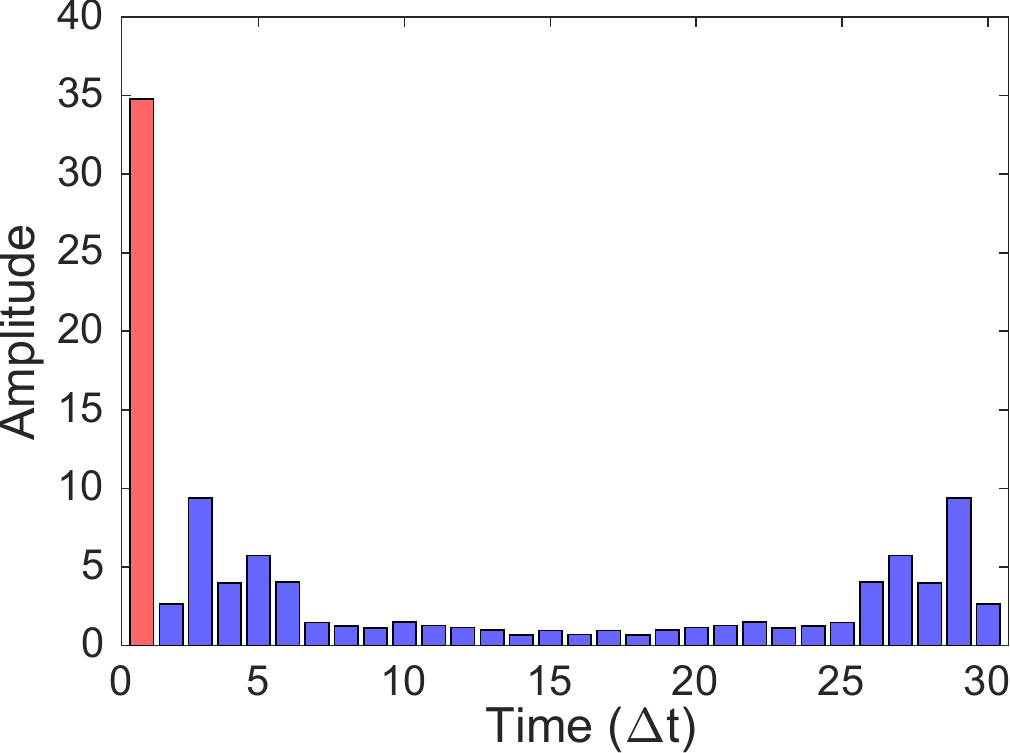}
}
\subfigure[CSI Frequency Domain.]{
\centering
\hspace{-0.1in}
\includegraphics[width=0.465\columnwidth]{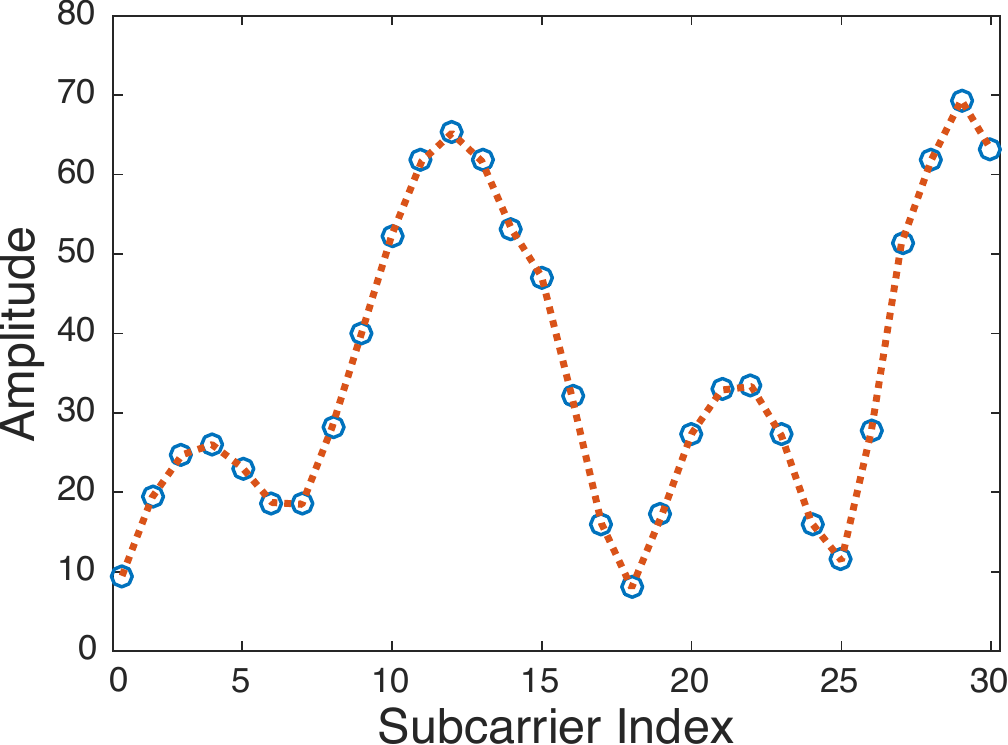}
}
\caption{Partial multipath effect mitigation. \textbf{Left:} in CSI time domain, we keep the item with the least time delay~(red box), and suppress~(dividing by 1000) remaining items~(blue box). \textbf{Right:} then we covert CSI time domain back to frequency domain. Above operations largely mitigate CSI compositions reflected from other paths, and focus on person body.}\label{fig:fft}
\end{figure}

\subsection{Feature Extraction}

 We retain CSI in all subcarriers since the absorption and reflection at different frequencies are necessarily to be involved in the ultimate features. Specifically, CSI at all subcarriers implies Wi-Fi signals present selective decline at different frequencies \cite{franceschetti2006wireless} over unique person body. After mitigating the impact of multi-path effect, we average CSI time series for 30 distinct features. Besides them, we leverage another 9 features to depict CSI frequency domain profile, one profile example shown in Fig.~\ref{fig:fft}~(Right). The features are the
(1) mean, (2) standard deviation, (3) median absolute deviation, (4) the mean absolute deviation, (5) interquartile range, (6)  root mean square, (7) skewness,  (8) kurtosis, and (9) entropy. The first 8 statistics values are common in time series mining~\cite{thomaz2015practical,reyes2016transition}, thus we only explain the entropy we proposed in WiPIN.




The entropy is used to describe the discrete degree of the CSI profile. Assume that the maximum and minimum values of CSI are $M$ and $m$, respectively. To calculate the entropy, we equally divide [m, M] into 10 bins, and count the number of CSI that fall in the $i$-th bin, i.e., $n_i$. Then we take $\frac{n_i}{30}$ as the probability that CSI falls in the $i$-th bin, donated as $p_i$. The entropy of CSI profile is computed via Equation.~\ref{eq:entropy}:
\begin{equation}\label{eq:entropy}
\boldsymbol{E} =- \sum_{i=1}^{10}p_i \log p_i.
\end{equation}
note that we define $p_i\log p_i=0$ if $p_i=0$.

After obtaining features, we segment collected CSI into training set and test set (segmentation detail will be descried in Section.~\ref{sec:evaluation}). Because metrics of these 39 features are not identical, we normalize them into [-1, +1] via following equation,
\begin{equation}\label{eq:scale}
x_i'= \frac{(2x_i -\max - \min)}{(\max - \min)},
\end{equation}
where $max$ and  $min$ are the maximum and minimum of the $i$-th features in the training set, respectively;  $x_i$ is the $i$-th feature of one training/test instance $x$, $x_i'$ is the corresponding normalized feature.

\subsection{Identity Matching}\label{sec:id}

\begin{figure}[tbp]\label{fig:recflow}
\centering
\includegraphics[width=0.9\linewidth]{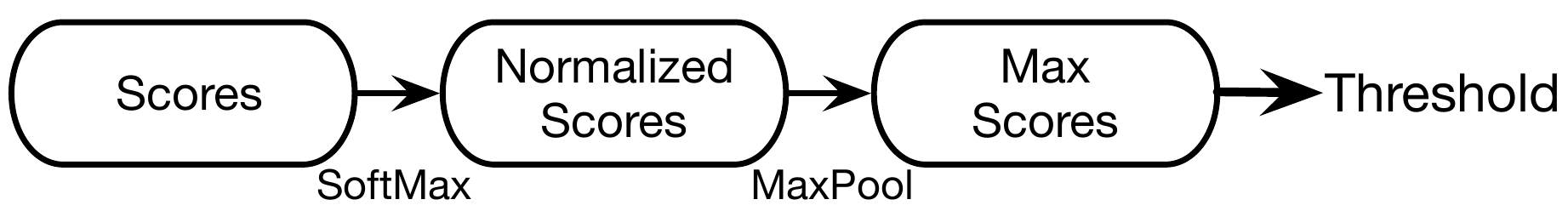}
\caption{Threshold learning process. We learn the identification threshold from the distribution of prediction scores. Then, those the maximal prediction score exceeding the threshold are considered as authenticated users. Once authenticated, index of maximal prediction score is the user ID prediction.}\label{fig:threshold}
\end{figure}

\subsubsection{Classifier Training}
Person ID classifiers are trained to identify the user ID. Using training set, we train person ID classifiers via a support vector machine (SVM) toolbox, i.e., LIBLINEAR~\cite{fan2008liblinear}. The classifiers are trained with the L2-regularized, L2-loss, primal, radial base function~(RBF) kernel, and their default hyper-parameters. Besides, we use one-against-all training strategy, which learns $N$ classifiers if the volume of users is $N$. The $N$ classifiers produce $N$  prediction scores on one instance. In this setting, the prediction ID is the index of classifier that outputs the highest score, formally as Equation.~\ref{eq:prediction}. 
\begin{equation}\label{eq:prediction}
i = \arg\underset{i \in{\{1,2,...,N}\}} \max  y_i.
\end{equation}
where $y_i$ is the score of classifiers on the $i$-th person.

\subsubsection{Threshold Learning}

According to Equation.~\ref{eq:prediction}, classifiers always make prediction from trained IDs~(aka authenticated IDs) even for illegal users, which makes WiPIN vulnerable. Thus we propose threshold learning algorithm to enable WiPIN to reject illegal users.
Recall that $N$ classifiers produce $N$ scores over trained users. With SoftMax function below, we normalize these $N$ scores into [0, 1]. 
\begin{equation} \label{eq:softmax}
y_i' = \frac{e^{y_i}}  {\sum_{i=1}^{N} e^{y_i}}.
\end{equation}
where $y_i$ is the same as Equation.~\ref{eq:prediction}; $y_i'$ is the normalized score. The normalized score can be interpreted as prediction confidence, e.g., $y_4'=0.6$ means that classifiers regard this person as the 4th user with 60\% confidence. 

\begin{figure}[tbp]
\centering
\includegraphics[width=0.85\linewidth]{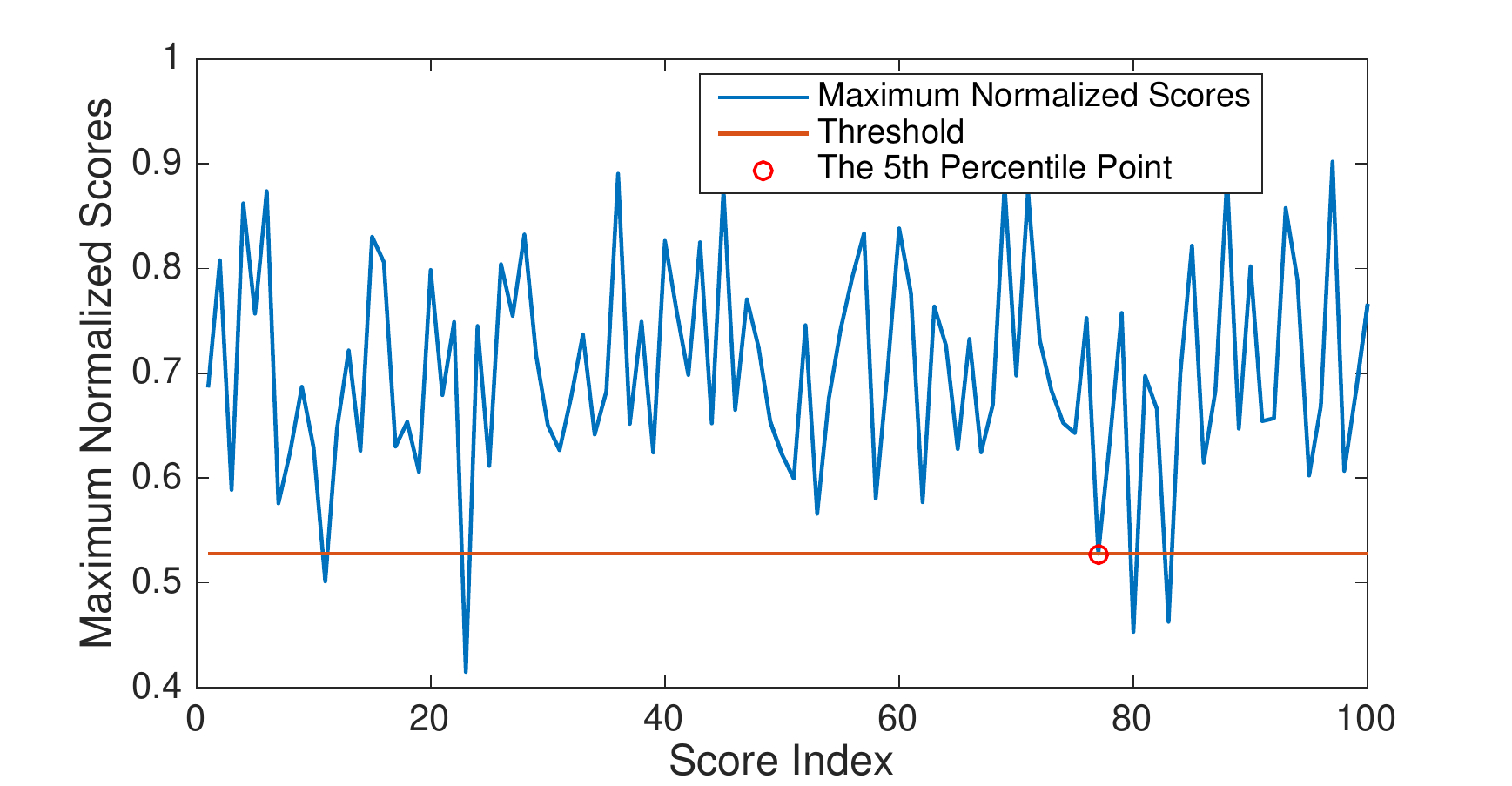}
\caption{100 examples of maximal normalized classifying scores. The value of the 5th percentile point~(red circle) as the rejecting threshold.}\label{fig:scores}
\vspace{-10pt}
\end{figure}

The threshold to identify illegal users is learned from the distribution of normalized scores. More precisely, 
for one training instance, we can compute its maximal normalized prediction score, denoted as $s$. Then we have the maximal prediction score set, $S$, computing from the whole training set. After removing mis-classified instances, we take the 5th percentile of these scores as the threshold. For example, we plot 100 maximal scores of 100 training instances in Fig.~\ref{fig:scores}. The 5th percentile is 0.5278, thus in this situation WiPIN regards a person as illegal user if her maximal normalized prediction score is lower than 0.5278. Otherwise, trained classifiers will output an ID prediction via Equation.~\ref{eq:prediction}.

%% file: tex/experiment.tex
\section{Experimental Evaluation}\label{sec:evaluation}

We use two Mini-PCs with Intel 5300 wireless NICs as the transmitter and receiver, respectively.
The frequency is set at 5GHz and the packet transmission rate is set to 500Hz. As illustrated in Fig.~\ref{fig:system}~(leftmost), the transmitter and receiver are placed on two cartons of 1.2$m$ high and  2.4$m$ apart. We recruit 30 subjects for our experiments, in which subjects are required to stand still at the mid-perpendicular of the Tx-Rx link.

\begin{figure}
    \centering
    \includegraphics[width=0.24\textwidth]{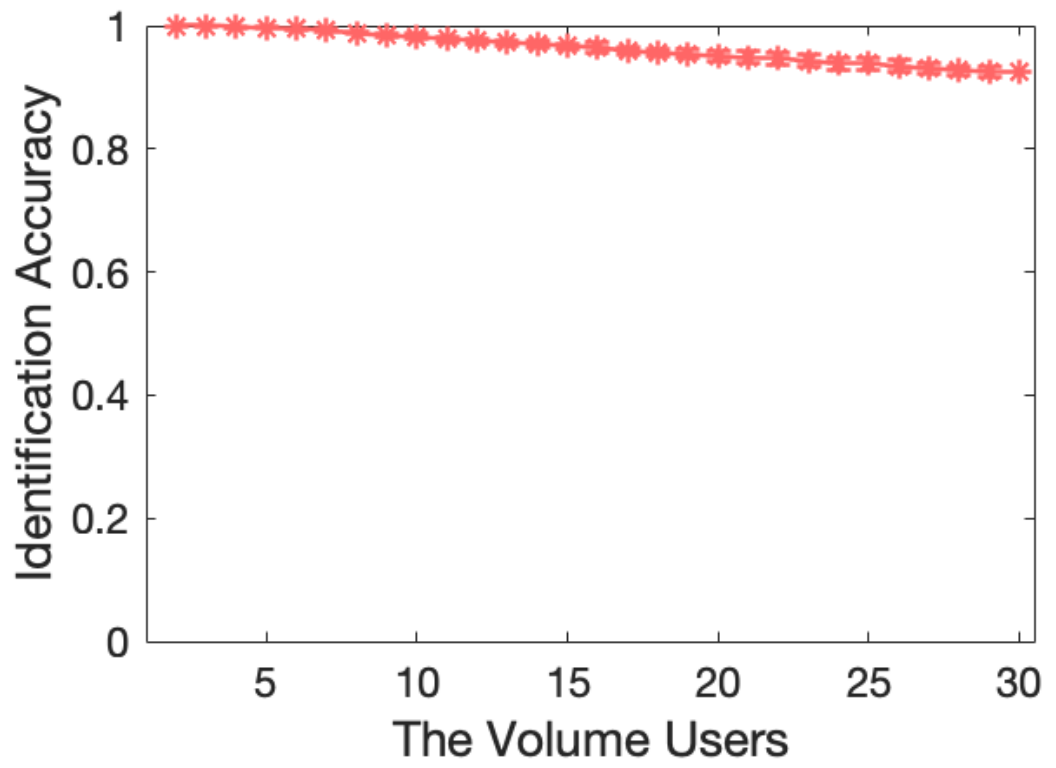}
    \includegraphics[width=0.24\textwidth]{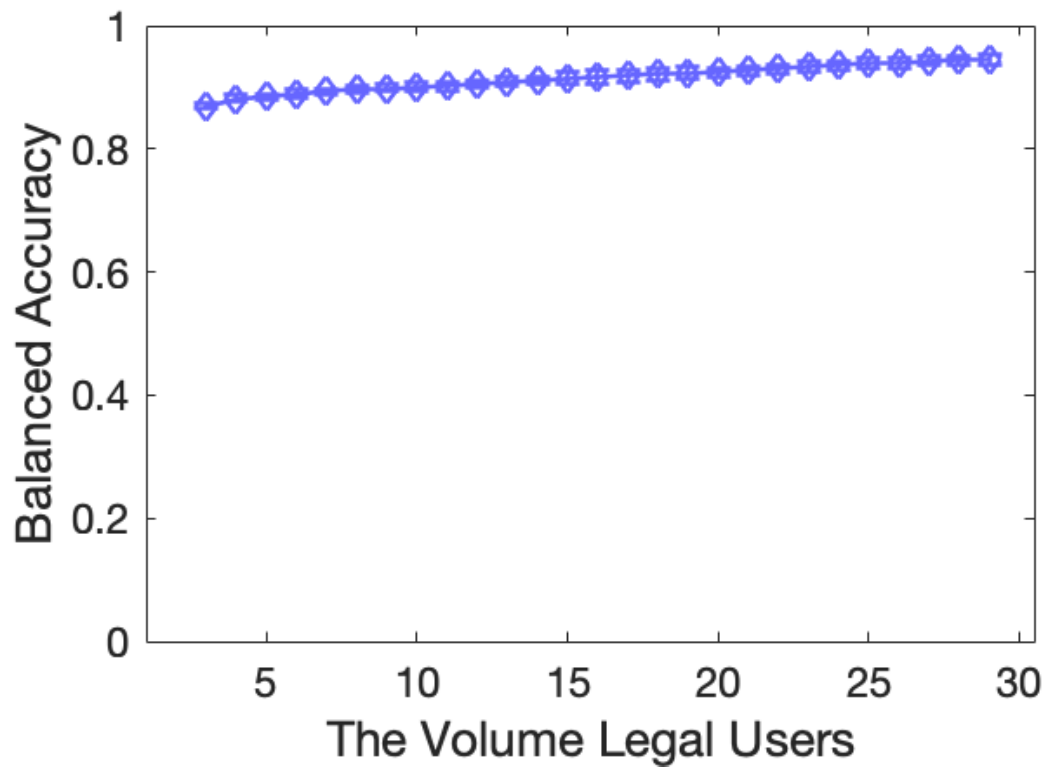}
    \caption{\textbf{Left.} When user volume increases from 2 to 30, identification accuracy gradually  decreases from 100\% to 92\%. \textbf{Right.} Evaluation on Threshold Learning Algorithm. We randomly select k (2 to 29) subjects as valid users to learn thresholds, meanwhile the remaining 30-k (28 to 1) users attack WiPIN. With more legal users and less attackers, balanced accuracy increases.}
    \label{fig:acc30}
    \vspace{-15pt}
\end{figure}

\begin{figure*}[tb]
\hspace{-0.1in}
\centering

 \hspace{-0.2in}
\begin{minipage}[t]{0.24\linewidth}
\centering
\includegraphics[width=0.93\textwidth]{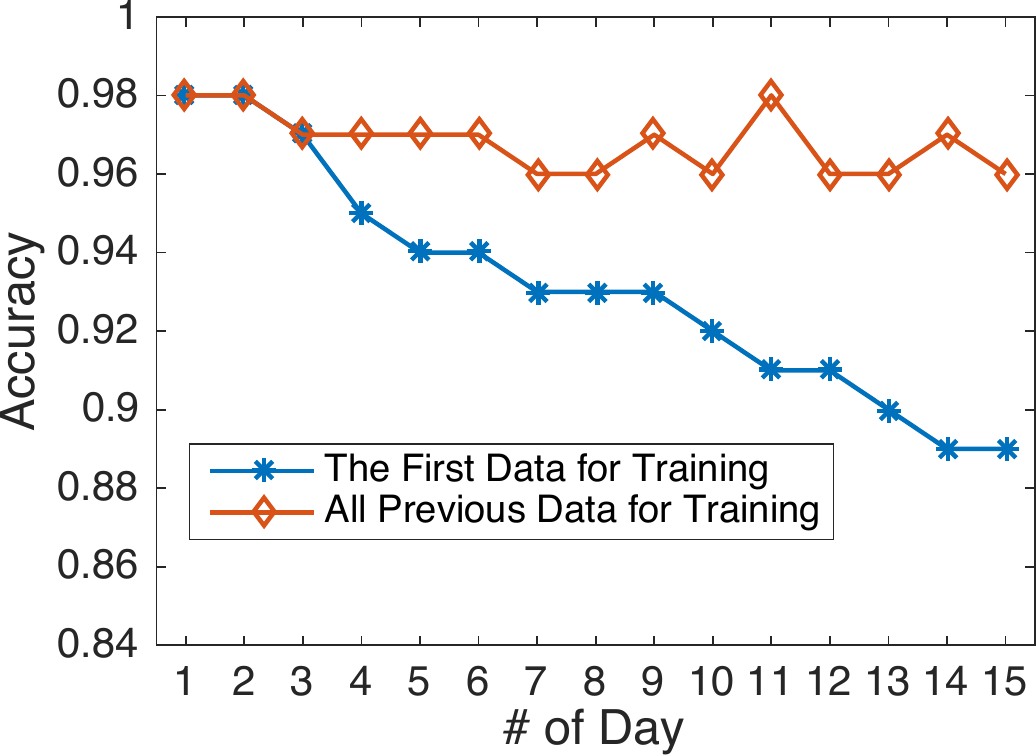}
\caption{Accuracy vs. Time.}
\label{fig:time_acc}
\end{minipage}
\begin{minipage}[t]{0.246\linewidth}
\centering
\includegraphics[width=0.97\textwidth]{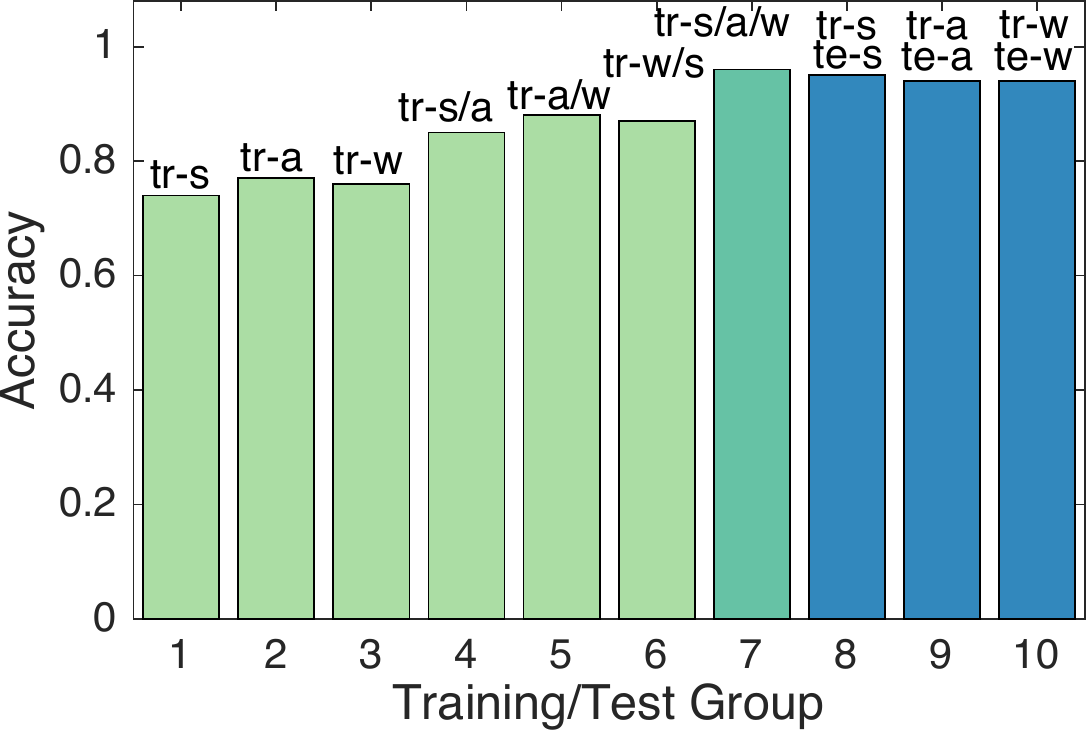}
\caption{Accuracy vs. Clothing.}
\label{fig:cloth_acc}
\end{minipage}
\begin{minipage}[t]{0.246\linewidth}
\centering
\includegraphics[width=0.975\textwidth]{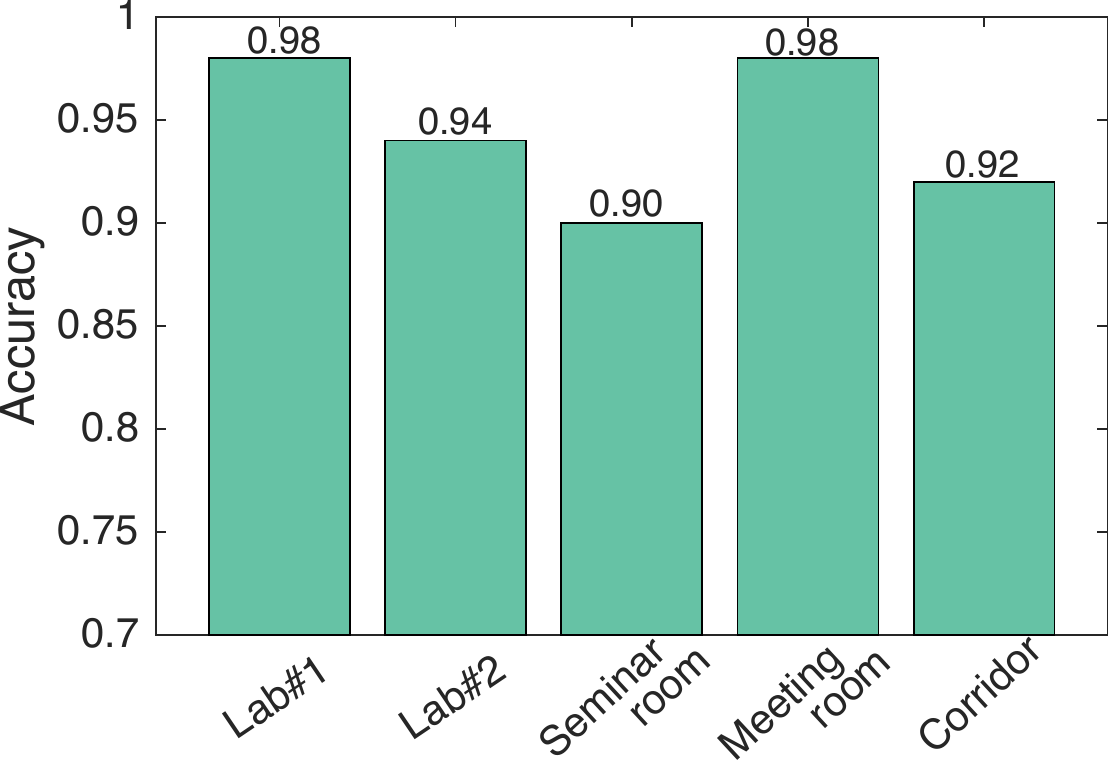}
\caption{Accuracy vs. Rooms.}
\label{fig:room_acc}
\end{minipage}
\begin{minipage}[t]{0.245\linewidth}
\centering
\includegraphics[width=0.9\textwidth]{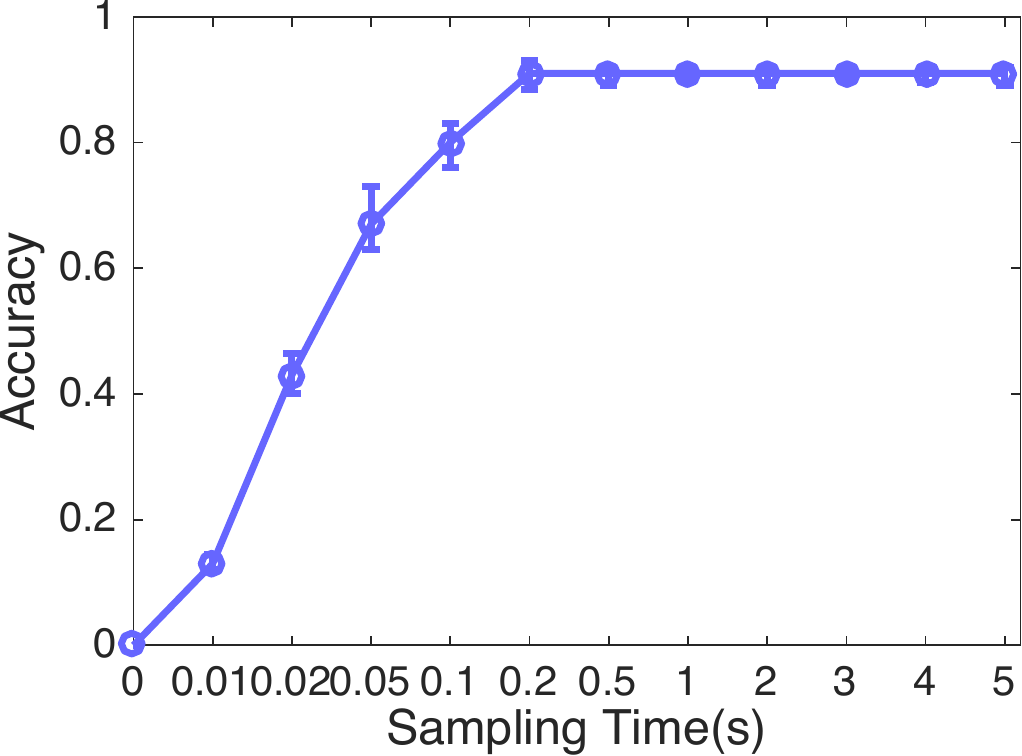}
\caption{Accuracy vs. Sampling Time.}
\label{fig:accu_samplingTime}
\end{minipage}
\end{figure*}

\begin{figure}[t]
\centering
\includegraphics[width=.9\columnwidth]{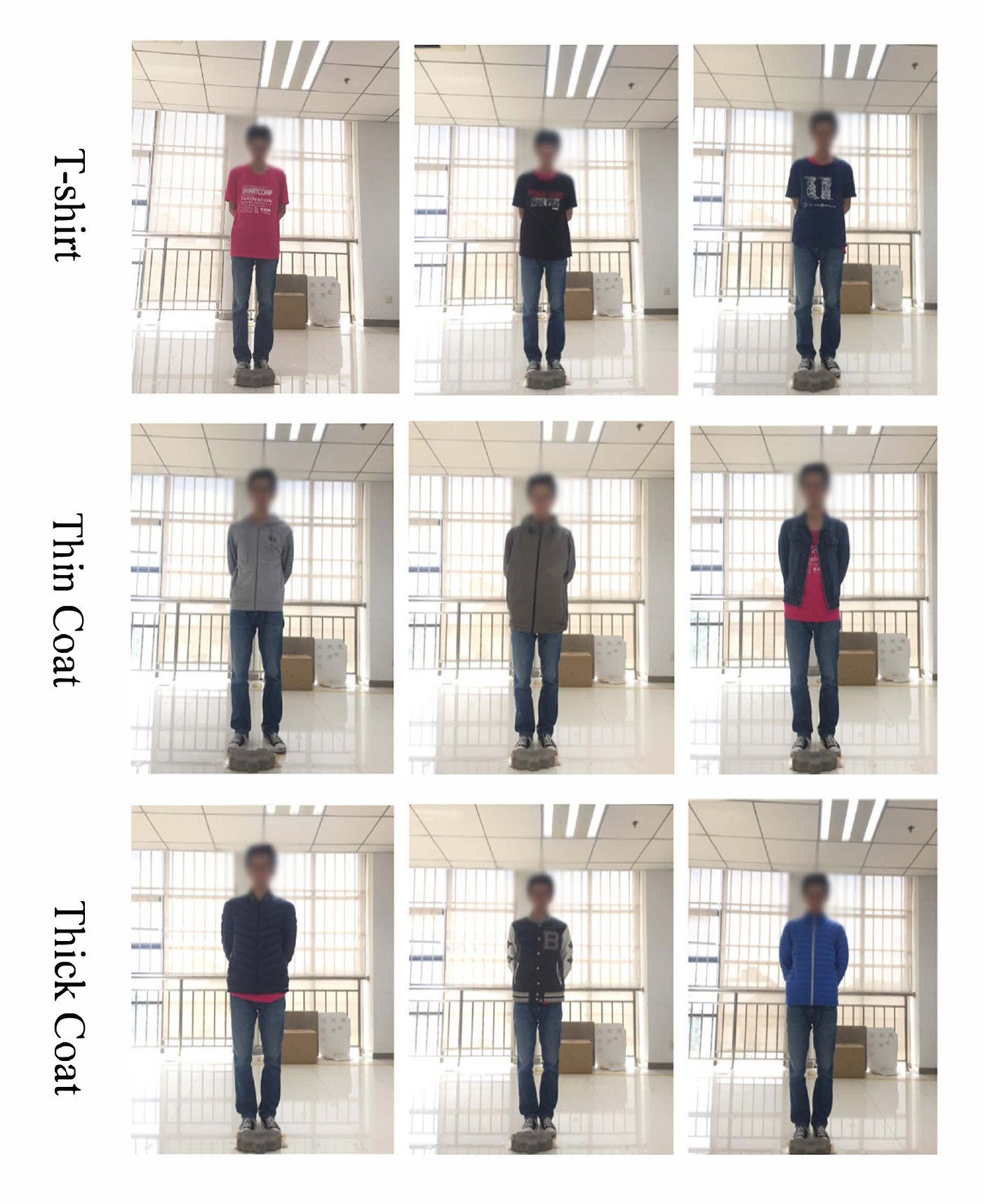}
\caption{Three Apparel Categories.}\label{fig:clothes}
\end{figure}

\subsection{Quantitative Results}\label{sec:accuracy}
For one subject, we record CSI time-serial data 30 times, each data lasting five seconds. We randomly select 20 out of 30 for training classifiers and learning defending thresholds, and the remaining 10 for the evaluation purpose. We perform the above data collection and segmentation on all 30 subjects.

\subsubsection{Performance of Identity Classifiers}


We would like to evaluate the accuracy of identity classifiers~(ratio of correctly identified in the test and the all test) with the user volume increasing. In particular, to evaluate the user volume of $k$, we randomly select $k$ subjects as users, apply their training data to train ID classifiers, and compute the test accuracy with their test data. Moreover, we do above randomly user selection, classifier training, and accuracy computing for 100 times. Thus for $k$, we have 100 accuracies, whose averages and quartiles are plotted in Fig.~\ref{fig:acc30}~(left). With this approach, we have an accuracy curve of ID classifiers. We see that WiPIN works pretty well in relatively small user volume, and gradually decreases to 92\% when all subjects are considered as users.

\subsubsection{Performance of Threshold Learning}
WiPIN can also reject illegal users that not seen before via threshold learning as described in Section.~\ref{sec:id}.
To test this, we select $k$ subjects as authenticated users to learn threshold, and apply the threshold on all subjects, where $k$ subjects are valid that should be classified as legal users~(true positive, TP), whereas the remaining $30-k$ users are illegal users that should be correctly rejected~(true negative, TN). We use the balanced accuracy~(BA = 0.5$\times$TP-Rate + 0.5$\times$TN-Rate) to evaluate the performance of the threshold learning algorithm. As similar scheme above, we compute BAs and quartiles of threshold learning with the legal user volume  increasing (illegal user volume decreasing), and show them in Fig.~\ref{fig:acc30}~(right).
We see the balanced accuracy increases from 0.87~(k=2) to 0.94~(k=29), maintaining at a high level.
In addition, we ascribe the the increment of BA to that the learning algorithm gain more knowledge in CSI of legal users to reject illegal attacks, i.e., higher TN-Rate, when trained with data that correspond to increasing legal users (with $k$ increasing).

\subsection{Evaluation on Robustness}

\subsubsection{CSI Stability vs. Time}
We recruit 10 subjects, record CSI, and prepare data as in Section.~\ref{sec:accuracy} over 15 consecutive days.
We apply 2 strategies to train ID classifiers to evaluate the stability of WiPIN.

Strategy 1: using the CSI data of the first day as the training set, and the data of other days as the test set.  This strategy simulates the scenario of no updates in WiPIN. Strategy 2: using all CSI data in the first $j$th days as the training set, and the data after the $j$th day as the test set. This strategy simulates the WiPIN-updating enabled scenario.

We plot results in Fig.~\ref{fig:time_acc}. We find that when using Strategy 1~(blue line), the accuracy of WiPIN decreases gradually. This result shows that time-varying does leave an impact on the feature stability, decreasing to 90\% after 10 days, implying the necessity of updating users' features. On the other hand, if we update WiPIN~(red line), it can keep high accuracy in certain period. With above analysis, we suggest that a proper updating period is about 10 days in this situation.

\subsubsection{Impact of Apparel Changing}

It is common that users change their apparels. In this evaluation, we roughly divide apparels into three categories, i.e., summer apparels (e.g., T-shirt), autumn/spring apparels (e.g., windbreaker), and winter apparels (e.g., down jacket).
We recruit 10 subjects and ask them to wear three categories of apparels, shown in Fig.~\ref{fig:clothes}. When one wears one clothes, we record corresponding CSI 15 times, where 10 out of 15 for training, the remaining 5 for test. Other settings in evaluation keep the same as in Section.~\ref{sec:accuracy}. We perform 10 cases of evaluation process.

Case 1-3. We alternately select CSI data collected when subjects wear one category of apparels as the training set, train WiPIN, and evaluate it via the data of all three categories. We call this One for All.

Case 4-6. We select the data of two categories of apparels as the training set, train WiPIN, and and evaluate it via the data of all categories. We call this Two for All.

Case 7. We select all training data, train WiPIN, and evaluate it via all test set. We call this All for All.

Case 8-10. We select the data of one category of apparels as the training set, train WiPIN, and predict by using the selected category of apparels as the test set. We call this One for itself.

We plot the average classification accuracy for each case in Fig.~\ref{fig:cloth_acc}.
The first six bars indicate that apparel changing to different categories has certain impacts on WiPIN. However, WiPIN can achieve at least 77\% accuracy when only utilizing one category apparel as the training set. On the other hand, bars of 7-10 in Fig.~\ref{fig:cloth_acc} demonstrate that WiPIN achieves an average accuracy of 94\% if the apparel category keep the same during training and testing. Above results mean that within a certain period, e.g., in summer or in wither, in which people generally do not change their apparels drastically, high performance can be guaranteed.

\subsubsection{Impact of Environment Noise} 
We choose five different places in our office building, including the places at an empty laboratory \#1, a crowded laboratory \#2, a crowded seminar room, a empty meeting room, and a narrow corridor, respectively. All these places are diverse in terms of surroundings, i.e., environment noise. We involve 10 subjects in this experiment. At each place, we collect 15 CSI series for every subjects, in which 10 are used as training data and the left as test data. Other settings keep the same as in Section.~\ref{sec:accuracy}. The results are shown in Fig.~\ref{fig:room_acc}.

We see that the average accuracy is about 94\%. Specifically, in the seminar room and corridor, WiPIN does not work as well as in laboratory \#1 and the meeting room. This is because the multi-path effect is much more stronger in the former places, making CSI components more complex and reducing the component weight that reflected from human body.

\begin{figure}[t]
\centering

\begin{minipage}[tb]{0.47\linewidth}
\centering
\includegraphics[width=0.95\textwidth]{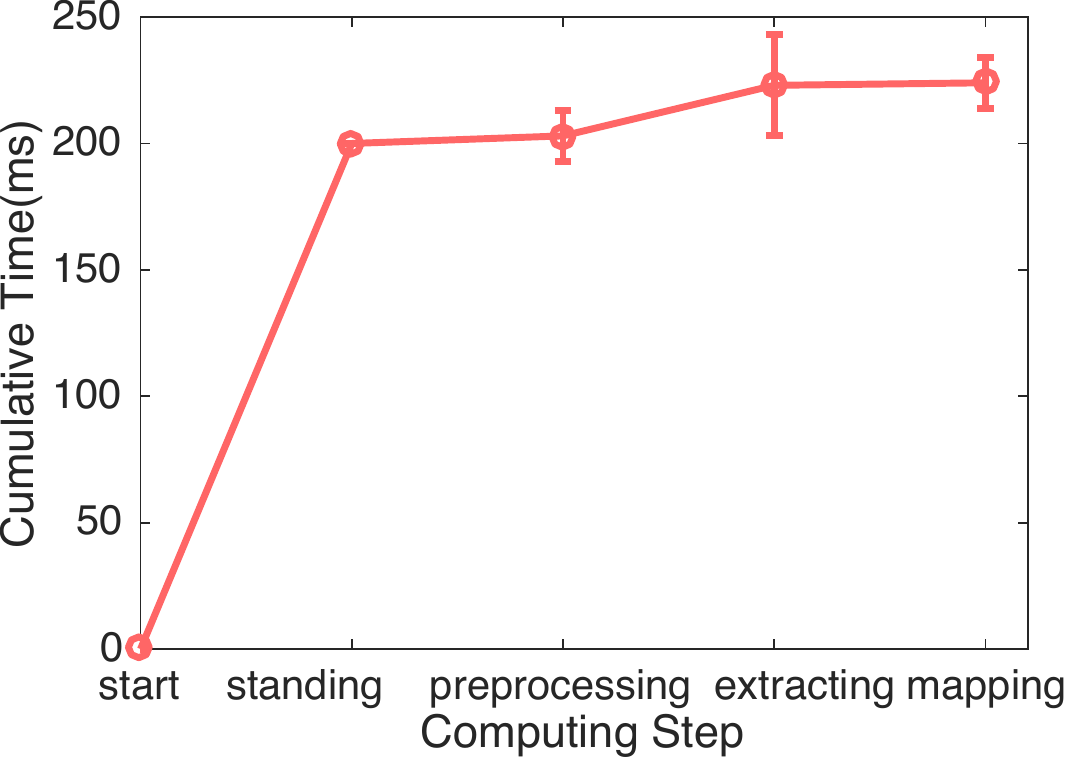}
\caption{Computation Overhead.}
\label{fig:overhead}
\end{minipage}
\begin{minipage}[tb]{0.47\linewidth}
\centering
\includegraphics[width=1\textwidth]{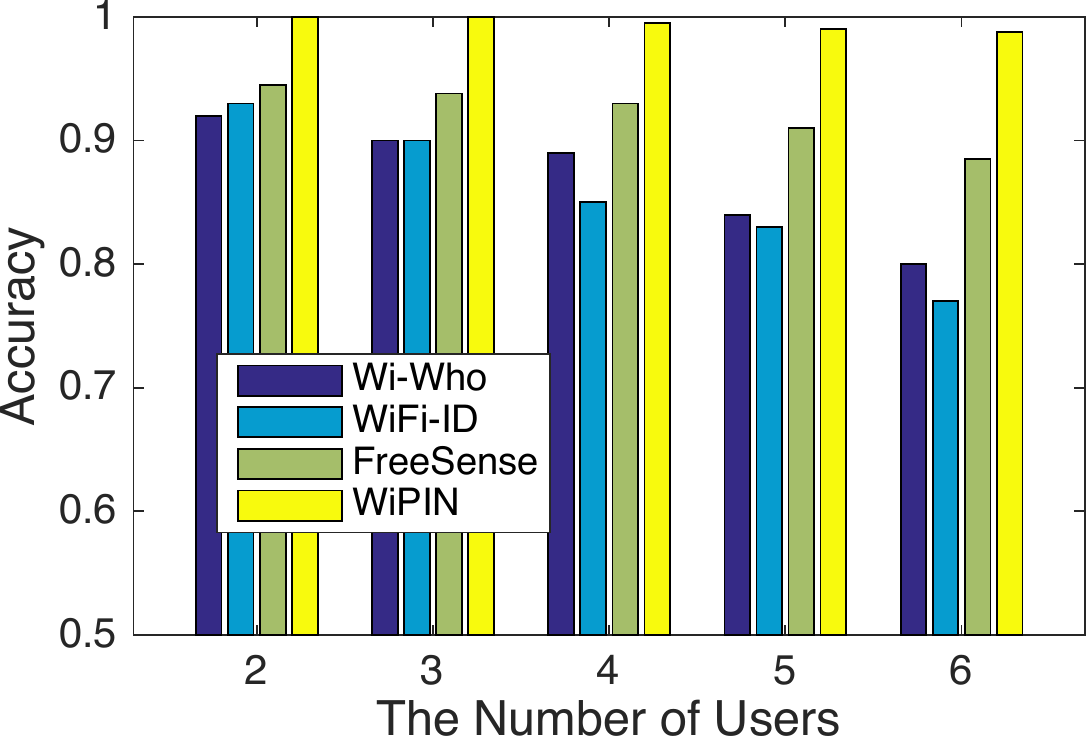}
\caption{Comparison to Prior Work.}
\label{fig:compare_acc}
\end{minipage}
\vspace{-10pt}
\end{figure}

\subsection{Computation Overhead}
We do computation overhead evaluation using data collected in Section.~\ref{sec:accuracy}. We adjust the sampling time from 0 to 5 seconds to 
determine the CSI sampling requirement~(user standing overhead) for a good identity matching performance.  Fig.~\ref{fig:accu_samplingTime} shows the results, where we see WiPIN can achieve 92\% mean accuracy while the time that user stands is more than 200$ms$.

Besides the user standing overhead, overhead in signal preprocessing, feature extraction, and identity mapping,
comprise computation time cost of WiPIN. According to Fig.~\ref{fig:accu_samplingTime}, we use CSI samples that collected in 200$ms$ to calculate the overhead.
The computation is done in a desktop PC~(with a CPU mode 2.7 GHz Intel Core i5, memory of 8GB DDR3) via Matlab R2015b. The cumulative computation overhead plus user standing overhead are shown in Fig.~\ref{fig:overhead}. indicating that WiPIN requires about 230$ms$ to identify a person, which is greatly time efficient. The low  overhead proves that WiPIN is applicable to real-time person identification applications.

\subsection{Comparison with Prior Work}
We compare WiPIN with three previous work, Wi-Who\cite{zeng2016wiwho}, WiFi-ID\cite{zhang2016wifi}, and FreeSense\cite{xin2016freesense}, in Fig.~\ref{fig:compare_acc}. As Fig.~\ref{fig:compare_acc} shows, WiPIN greatly outperforms those three approaches in terms of accuracy.
In addition, these works are all operation-based approaches, requiring the user to walk 2$m$-6$m$, which is inconvenient and raises barriers for real-world applications. 

%% file: tex/related.tex
\section{Related Work}

Passive person identification that mainly utilizes person behavior patterns, such as when typing~\cite{de2012touch, shahzad2013secure} and breathing~\cite{chauhan2017breathprint}, has proven popular. It is hard to fool these system because the attack needs extremely vivid imitation on user behaviors. Among past work, applying Wi-Fi signals to detect person walking pattern (aka gait) is one leading schema in wireless security and privacy community, such as WiFiU~\cite{wang2016gait}, WiWho~\cite{zeng2016wiwho},  WiFi-ID~\cite{zhang2016wifi}, etc.
In gait-based work, to get authenticated, users must walk along the pre-defined path several meters, which is labor intensive and time-consuming, thus making it unpractical for use. In addition, an attacker can record a video when users walking, then practice to walk with a similar gait as the users to be a fake. 

Compared to gait-based work, WiPIN requires user to do no operation, but to stand for more than 200$ms$, which is user-friendly, time efficient, and meets requirements of real-time use scenarios. Moreover, WiPIN captures whole body information that is with high resilient to attacks.